\documentclass[aps,prb, twocolumn]{revtex4}
\usepackage{epsf}
\usepackage{bm}
\usepackage[utf8]{inputenc}
\usepackage{textcomp}
\usepackage[pdftex]{graphicx}
\graphicspath{{./img/}}
\newcommand\bra{\left\langle}
\newcommand\ket{\right\rangle}
\newcommand\mb{\mathbf}
\renewcommand\Im{\mathop{\rm Im}\nolimits}
\renewcommand\Re{\mathop{\rm Re}\nolimits}

\begin{document}
\title{Optical properties of a disordered metallic film: local vs. collective phenomena}
\author{A.A. Osipov$^1$, A.N. Rubtsov$^{1,2}$}
\affiliation{$^1$Department of Physics, Moscow State University, 119991 Moscow, Russia\\
$^2$Russian Quantum Center, 100 Novaya St., Skolkovo, 143025 Moscow, Russia}
\date{\today}
\begin{abstract}
We apply the dual-variables approach to the problem of the optical response of a disordered film of metal particles with dipole-dipole interaction. 
Long range dipole-dipole interaction makes the effect of spatial correlations significant, so that dual-variables technique provides a desirable improvement of the coherent-potential results. 
It is shown that the effect of nonlocality is more pronounced for a medium-range concentration of the particles. 
The result is compared with the nonlocal cluster approach. 
The short-range correlations accounted in the cluster method reveal themselves in the spectral properties of the response, whereas long-range phenomena kept in the dual technique are more pronounced in the $k$-dependence of the Green's function. 
\end{abstract}
\maketitle
\section{Introduction}
\label{sec:Intro}
Disordered systems of a various physical nature with different typical length-scales play an important role in condensed matter physics.
Among the systems with a disordered atomic configuration we find various alloys, amorphous semiconductors and diluted magnetics. Disordering at a nanoscale level is typical for arrays of semiconductor quantum dots and metallic heterostructures. 
Depending on the length-scale, the disorder reveals itself in different experimental observables. 
For systems with an atomic length-scale disorder, the electronic properties are the most interesting. 
For random arrays of nanometer-size and larger particles, the study of their optical response is a remarkably active field of research.

There is a number of common approaches to describe the different phenomena in disordered systems. 
In particular, the approach of the coherent potential approximation (CPA) has been used for the description of the electronic system in the substitutional alloy model \cite{Soven:1967}, but it has also been applied in the description of the optical properties of disordered nanostructures. Nevertheless, there is a considerable difference in the type of the interaction (short-range hopping for electrons and long-range Coulomb interaction in the optics of nanostructures). 
One should also note that optics offers much larger set of observables; for example, generation of high order optical harmonics allows experimentalists to study many-particle Green's functions of photons.
Nowadays, the problem of theoretically describing and calculating the optical properties of such systems have gained interest because of progress in the technology of  preparing novel materials \cite{Heilmann.mono}.
For a theoretician working in the field of electronic structure, a description of the optical response provides an additional possibility to validate existing calculation schemes and it can be a good playground for the development of new methods.

In this paper we consider a model describing disordered films of metallic particles.
Apart from the remarkable progress in theoretical methods as well as in experimental techniques, a relatively small number of theoretical predictions have been quantitatively compared with experimental data so far. This is because of the difficulty in preparing a well-controlled sample exactly corresponding to the system under theoretical consideration (a lattice gas, for instance). 
On the other hand, the significant increase of computing power in the last few decades makes it possible to undertake direct numerical simulations of such systems. This allows to check the validity of theories in the most direct way. 
Such calculations are one of the subjects of this paper.  

Another, and more important point is an analysis of the nonlocal physics beyond the CPA. For this purpose, we 
study a system where the CPA and its extensions are easy to formulate: the  so-called lattice gas being a periodic lattice with metallic particles randomly placed at nodes \cite{Persson:1983}. 
The CPA corresponds to a substitution of the partially filled lattice with a regular one, having particles with a self-consistent permittivity at each node \cite{Persson:1983}, \cite{Persson:1983a}, \cite{Barrera:1991}. 
Physically this means that fluctuations of electromagnetic field at different sites are assumed as uncorrelated (that is, fluctuations are local).

There are a few arguments for physics beyond the single-site CPA. 
Firstly, in some systems there are correlations due to grouping of the disordered particles. For example, a fractal nature of clusters is assumed in a series of works \cite{Hui:1986}, \cite{Chang:1992}, \cite{Markel:1996}, \cite{Markel:2004}. 
The idea that in a set of randomly distributed particles some specific configurations of particles placed close to each other gives a significant contribution to the total polarizability is applied in \cite{Kochergin:2007}.
In the present paper, we stick to another case when the particles are distributed independently, but spatial correlations appear in the electromagnetic field at different particles. 

For the description of nonlocal field correlations, there are strong insights from the theory of electronic structure in the Anderson model, since the mathematical formulations of both problems almost coincide.
While the CPA was successfully used in a number of calculations for the electronic properties of metallic alloys \textit{ab initio}, certain peculiarities in the electronic density of states are attributed to nonlocal correlations. 
A trivial generalization of the single-site CPA is to introduce a ``supercell'' containing several atoms, and to suppose that correlations are located in such a real-space cluster.
However, such an approach clearly violates the translational invariance of the lattice and it also does not provide a significant improvement to single-site theory \cite{Tsukada:1972}, \cite{Ducastelle:1974}.
It is much better to formulate a theory operating in quasimomentum space, as is done in the dynamical cluster approximation (DCA)\cite{Jarrell:2000}.
The idea of the DCA was modified and applied to the Anderson model of disordered systems\cite{Jarrell:2001}, and it is known as the nonlocal coherent potential approximation (NLCPA).
It is shown that this method can be reformulated as a unique reciprocal-space theory of disorder\cite{Rowlands:2008}. 


An alternative to cluster methods is the so-called dual-variable approach (DVA).
This method was originally introduced for classical lattice ensembles \cite{Rubtsov:2002} and correlated fermions \cite{Rubtsov:2008}, as an extension of the effective-medium approximations.
Recently, the DVA have been successfully applied for the Anderson localization problem \cite{Terletska:2012} for disordered systems. It is worth to note that in all mentioned papers an introduction of dual variables is based on the Hubbard-Stratonovich transformation in the expression for the partition function. 
In the present paper we show that for disordered system, at least at the basic level, neither the Hubbard-Stratonovich transformation nor replica method are necessary for the construction of the DVA. 
The change of variables and further construction of the diagram series can be done directly in the expression for the linear response of the system.  

In the ``Definitions'', we describe details of the system under consideration and discuss methods used for study of the response function of those systems. 
The ``Calculations'' section is devoted to computational details, and in the ``Results'' we compare the numerical simulations and the results of analytical schemes. 
\section{Definitions}
\label{sec:Def}
We consider the optical response of a planar film containing spherical metallic particles. 
The size of particles and the lattice constant are assumed negligible in comparison with the wavelength of the incident radiation.
Also for simplicity, we consider only the case of the electrical field being normal to the film plane.
The general case can be considered similarly \cite{Persson:1983}.
Particles are placed at the nodes of a regular square lattice (arbitrary lattice symmetry could be assumed), but their susceptibilities are random uncorrelated quantities.
In the dipole approximation, this system is described by the system of equations for the dipole moments $d_i$:
\begin{equation}
\label{dipole-regular-lattice}
	d_{i} = \alpha_i\left(E_{i}^{ext} + \sum_{j}C_{ij}\cdot d_{j}\right),
\end{equation}
where $\alpha_i$ are (random) susceptibilities of the particles, and $C_{ij}$ describes their dipole-dipole interaction.
Since we stick to the Rayleigh limit, $C_{ij, i\neq j} = - \frac{C_{0}}{r^{3}_{ij}}$ ($C_{0}$ is the interaction constant, $C_{ii} = 0$ to exclude a self-action). 
Note that in the optical problems susceptibility has a finite imaginary part, whereas in problem of the electrons localization an infinitesimal displacement of poles from the real axis is assumed. 
In this paper, we are interested in the average value of the dipole moment, $\langle d \rangle$.
Using the ergodic hypothesis one can equally average over the sites of an infinite system or over the realizations.
A distribution function of the susceptibility $\alpha$ should be specified.
We assume a binary distribution, so that $\alpha_i$ takes the value $\alpha^+$ with a probability $p$, and $\alpha^-$ with a probability $(1-p)$.
There is an important special case of $\alpha^{-} = 0$, which corresponds to a partially filled lattice (lattice gas) with a filling factor $p$.
It also should be pointed that $\alpha$ has a spectral dependence $\alpha = \alpha(\omega)$, where $\omega$ is the frequency of the incident field.

The equations (\ref{dipole-regular-lattice}) can  be represented in a matrix form:
\begin{equation}
	\label{vec-dipole-regular-lattice}
	\mb{d} = \alpha\left(\mb{E}^{ext} + C\cdot\mb{d}\right).
\end{equation}
In real-space, the matrix $\alpha$ is diagonal, with random uncorrelated elements. 
The regular matrix $C$ becomes diagonal after the Fourier transform.
The solution of (\ref{vec-dipole-regular-lattice}) after averaging over system configurations gives the average dipole moment of the whole system and can be represented as follows:
\begin{equation}
	\label{GF-def}
	\bra\mb{d}\ket = \mb{E}^{ext}\bra\frac{1}{\alpha^{-1} - C}\ket \equiv \mb{E}^{ext}G,
\end{equation}
where $G$ is the response function.
The later is diagonal in quasimomemtum space ($k$-space), due to the translational invariance of the averaged system.
Mention here that $\Re G_{\mb{k}=0}$ function represents the average dipole moment of the whole system, whereas $\frac{1}{N}\Im G_{\mb{k}}$ determines the density of state (DoS) of the system.
Such phenomena as localization, optical scattering and high harmonic generation are determined by higher order Green's functions.
For a weak disorder the standard diagrammatic expansion is valid, in general case a more sophisticated consideration is needed.
\begin{figure}[h]
\includegraphics[scale=0.05]{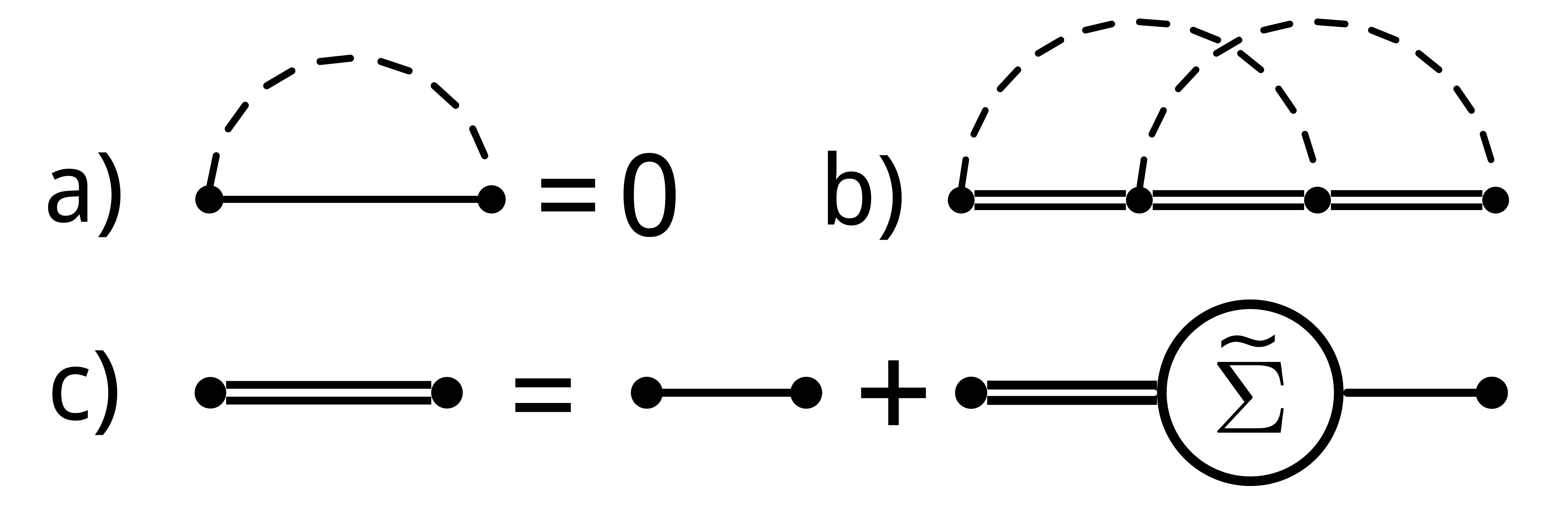}
\caption{Diagrammatic notations}
\label{fig:Diag}
\end{figure}

\subsection{The effective-medium concept}
The most trivial effective-medium approach to calculation of the response of a disordered system is to replace all the dipole moments with an average value, so that (\ref{dipole-regular-lattice}) become:
\begin{equation}
	\bra d\ket = \bra\alpha\ket \left(E_{i}^{ext} + \sum_{j}C_{ij}\cdot \bra d\ket\right).
\end{equation}
For the system under consideration this approach is the so-called virtual crystal approximation (VCA).

The Fourier-transform gives the solution in reciprocal space $d_{\mb{k}}^{\mathrm{VCA}} = G_{\mb{k}}^{\mathrm{VCA}} E_{\mb{k}}$ with 
\begin{equation}
	G_{\mb{k}}^{\mathrm{VCA}} = \frac{1}{\bra\alpha\ket^{-1} - C_{\mb{k}}}.
\end{equation}
One can observe that such a mean-field approximation corresponds to neglecting the correlations between the susceptibility of a particle and the acting field;
mathematically this is the decoupling 
$\bra\alpha_i  \sum_{j}C_{ij}\cdot d_{j}\ket\to \bra\alpha_i\ket  \sum_{j}C_{ij}\cdot \bra d_{j}\ket$.
Strictly speaking, this is not true even for a single fluctuating site placed in a regular medium.
Indeed, the dipole moment results in a polarization of the medium, and that polarization contributes to the field acting on the site. 
This is similar to the effect of the so-called electrostatic image, that acts on a dipole placed near a substrate.
The larger the dipole moment is, the larger the image field becomes. 
So, the susceptibility and the field become correlated, and the mean-field decoupling breaks down.

Such ``trivial'' correlations are completely taken into account within the CPA.
In this approximation, a hybridized single-site problem is considered, so that instead of $\bra\alpha\ket$ one calculates the average:
\begin{equation}\label{CPAimp}
	g\equiv\left<\frac{1}{\alpha^{-1}-\Delta}\right>,
\end{equation}
where $\Delta$ represents the regular self-influence of the node due to the environment (bath).
The average response is now given by the Green's function
\begin{equation}
\label{GF-CPA}
	G_{\mb{k}}^{\mathrm{CPA}}=\frac{1}{g^{-1} + \Delta - C_{\mb{k}}}.
\end{equation}
As a last step of the CPA formulation, one requires that the single-site average (\ref{CPAimp}) mimics the local part of the lattice Green's function,
\begin{equation}
\label{CPA-SCC}
	g = \frac{1}{N}\sum_{\mb{k}} G_{\mb{k}}.
\end{equation}
This equation allows to find self-consistently the hybridization $\Delta$.

\subsection{The nonlocal CPA}
The natural way to further develop an effective medium theory, is the construction of a self-consistent scheme, which takes into account nonlocal correlations, using the self-consistency condition for a set of sites (cluster). 
The main disadvantage of such methods is the violation of the translational invariance of the underlaying lattice as a result of the embedding of a real-space cluster into the medium. 
In the NLCPA self-consistent clusters with the periodical boundary conditions are constructed in the reciprocal space, and the effective medium in this method preserves the translational invariance of the initial lattice.
It is constructive to represent the NLCPA effective medium in coordinate space not as a set of site clusters, but as a set of interacting sublattices.
In the reciprocal space this is equivalent to dividing the Brillouin zone (BZ) of the initial lattice into $N_c$ tiles. 
In contrast with the single site approximation for the self-energy in the CPA, the exact self-energy in the NLCPA is approximated as a step function with the number of steps in reciprocal space, corresponding to the number of considered sub-lattices:
\begin{equation}
	\Sigma(\mb{k})\simeq\Sigma(\mb{K}_n).
\end{equation}
This assumption accounts for the nonlocal correlations described by off-diagonal elements of the self-energy only for sites located on different sublattices.

The next approximation made in the NLCPA is setting the phase factors in the Fourier transformation to unity for all $\mb{k} \neq \mb{K}_n$.
Physically this means that each site of the cluster interacts with the medium irrespective of its position in cluster.

The corresponding Green's function is represented by coarse-grained propagators $g_{\mb{K}}$ ($\mb{K}$ is taken from the set of $\mb{K}_{n}$):
\begin{equation}
\label{CG-GF}
	g_{\mb{K}}\equiv\frac{N_c}{N}\sum_{\tilde{\mb{k}}}G(\mb{K} + \tilde{\mb{k}}),
\end{equation}
where the $\tilde{\mb{k}}$ - summations run over the momenta of the $n$-th cell, $N$ is the number of nodes in the lattice, and $N_c$ is the number of tiles.
The explicit representation of the coarse-grained Green function in reciprocal space has a standard form:
\begin{equation}
\label{CG-GF-NLCPA}
	g_{\mb{K}} = \frac{N_c}{N}\sum_{\tilde\mb{k}} \frac{1}{\bra\alpha^{-1}\ket - C_{\mb{K}+ \tilde{\mb{k}}} - \Sigma_{\mb{K}}}.
\end{equation}
It is necessary to introduce the cluster-excluded propagator (also called the cavity Green's function) $\mathcal{G}_\mb{K}$ to avoid the occurrence of a cluster self-action:
\begin{equation}
\label{CE-GF}
	\mathcal{G}^{\mathrm{NLCPA}}_{\mb{K}} = \frac{1}{g^{-1}_{\mb{K}} + {\Sigma}_{\mb{K}}}.
\end{equation}
The self-consistent scheme of this approach is based on a requirement that the coarse-grained Green function equals the Green's function for the cluster placed in a corresponding cavity:
\begin{equation}
\label{GF-NLCPA}
	G^{\mathrm{NLCPA}} = \bra\frac{1}{\alpha^{-1} + \mathcal{G}^{-1} - \bra\alpha^{-1}\ket}\ket_{\mathrm{cls}},
\end{equation}
where the average is taken over the cluster configurations.
In a practical calculation, the right-hand side of the expression (\ref{GF-NLCPA}) is calculated in the real space, with a subsequent Fourier transform.
For a small cluster and a binary distribution of $\alpha_i$, the cluster Green's function can be found analytically.
Formulas (\ref{CG-GF-NLCPA}-\ref{GF-NLCPA}) form a closed set of the NLCPA equations. 

To show similarity between the CPA and the NLCPA, one can rewrite (\ref{GF-NLCPA}) as follows:
\begin{equation}
	G_{\mb{K}} = \bra\frac{1}{\alpha^{-1} - \Delta_{\mb{K}}}\ket_{\mathrm{cls}},
\end{equation}
where $\Delta_{\mb{K}} \equiv \bra\alpha^{-1}\ket- g^{-1}_{\mb{K}} - \Sigma_{\mb{K}}.$
In this notation (\ref{CG-GF-NLCPA}) appears to be an extension of (\ref{CPA-SCC}):
\begin{equation}
	g_{\mb{K}} = \frac{N_c}{N}\sum_{\tilde{\mb{k}}}\frac{1}{g_{\mb{K}}^{-1} + \Delta_{\mb{K}} - C_{\mb{K} + \tilde\mb{k}}}.
\end{equation}
Clearly, the NLCPA passes into the CPA for $N_c = 1$.
In the limit of $N_c \rightarrow N$ the results of the NLCPA become exact. 

The CPA contribution to the self-energy part consists of only the members corresponding to scattering at a single node, and consequently the self-energy is $\mb{k}$-independent: $\Sigma_{\mb{k}}\equiv\Sigma$.
Because of the assumption about the approximation for the exact self-energy, the NLCPA takes into account (in an approximate way) nonlocal correlations at the length-scale comparable with a cluster size.
\subsection{Dual variables approach}
\label{sec:DVA}
We start the discussion of the dual variables method by formally introducing the following expression: 
\begin{equation}
  \label{eq:DV-intro}
	M = (C - \Delta)^{-1} - (\alpha^{-1} - \Delta)^{-1},
\end{equation}
where $\Delta$ is a diagonal, regular matrix. 
Inversion and subsequent averaging of the (\ref{eq:DV-intro}) gives, with an account of eq. (\ref{GF-def}), the following form for $\tilde{G}\equiv\left\langle M^{-1}\right\rangle$:
\begin{equation}
  \label{GGdual}
	 \tilde{G} = (C-\Delta)  G  (C-\Delta)+(C-\Delta),
\end{equation}
so to obtain $G$ one can calculate the dual Green's function $\tilde{G}$.
We may assume that it can be represented as an expansion:
\begin{equation}
	\tilde{G} = \bra\frac{1}{R^{-1} + Y}\ket = R - \bra RYR\ket + \cdots,  
\label{M-exp}
\end{equation}
where $R=\bra M\ket^{-1}$ and $Y\equiv M - \bra M \ket$. Straightforward calculation gives
\begin{equation}
	R = g^{-2}\left(\frac{1}{g^{-1}+\Delta-C}-g\right),  
\label{defR}
\end{equation}
and 
\begin{equation}
Y=g - (\alpha^{-1} - \Delta)^{-1}.
\label{defY}
\end{equation}
Diagrammatic techniques can be used for representation and further work with this expansion.
By construction, we require such $\Delta$ that 
\begin{equation}
	\bra Y\tilde{G} Y\ket = 0,
\label{SCC-DVA}
\end{equation}
at each order of the expansion (self-consistent condition).
This condition in diagrammatic form reads (Fig.~\ref{fig:Diag}a),
where the solid line represents the dual propagator, and the dashed line is corresponds to influence of the disorder.
Since $Y\neq 0$, in reciprocal space this requirement is represented as follows:
\begin{equation}
	\sum_{\mb{k}}\frac{1}{(C_{\mb{k}} - \Delta)^{-1} - \bra(\alpha^{-1} - \Delta)^{-1}\ket} = 0.
\label{SCC-RS}
\end{equation}
At the zeroth order of the expansion (\ref{M-exp}) $\tilde{G}=R$ and it immediately follows from (\ref{defR}) that:
\begin{equation}
	g = \frac{1}{N}\sum_{\mb{k}}\frac{1}{g^{-1} + \Delta - C_{\mb{k}}}.
\label{SCC-explicit}
\end{equation}
It is clear that later is the same as the self-consistent condition of the CPA (\ref{CPA-SCC}). Moreover, 
the relation (\ref{GGdual}) where $\tilde{G}$ is substituted with $R$, gives exactly the CPA formula (\ref{GF-CPA}) for $G$.
In other words, the zeroth order of this approach is the single-site self-consistent effective medium approach, equal to the CPA.
This allows to justify the expansion (\ref{M-exp}). Since the CPA is known to take into account all local correlations, it means that in the present approach the local part of correlations is taken into account by the special choice of the matrix $M$ with a proper $\Delta$. Consequently, the series (\ref{M-exp})  expands around this result, taking into account the nonlocality of correlations. In this context, it is instructive to note that $R$ is proportional to the CPA Green's function (\ref{GF-CPA}) with the local part excluded, so that dual corrections indeed are essentially nonlocal.
It follows from the above statement that the DVA corrections are small near the limit cases where the CPA becomes exact (the cases of a small disorder, small coupling, and a large coordination number). For each limit case 
a corresponding formal small parameter can be pointed out. For a general situation,
there is a physical justification of the DVA: one should expect that the theory behaves well if the CPA 
provides a good starting point.

One can observe not only the ideology of the DVA resembles the dual-fermion approach\cite{Rubtsov:2008}, but also formal 
expressions (\ref{GGdual}) and  (\ref{SCC-DVA}) coincide with corresponding expressions for the dual fermions, although the Hubbard-Stratonovich transformation was not used in the presented approach for a disordered system. 

Let us now consider the leading-order correction to the CPA, as it appears from the expansion (\ref{M-exp}).
As long as $\bra Y\ket = 0$, all odd members of this expansion vanish.
The second term of the (\ref{M-exp}) is also equal to zero, since we require (\ref{SCC-DVA}) by construction of the method.
It now follows that the next non-vanishing term of the (\ref{M-exp}) is of the fourth order in $Y$.
We can use standard techniques to achieve the following representation of this approximation in the form of the Dyson equation:
\begin{equation}
\label{Dyson-dual}
	\tilde G^{-1} = R^{-1} -\tilde \Sigma,
\end{equation}
where $\tilde \Sigma$ is the dual self-energy part. 
The diagrammatic representation of this equation has the usual form shown in Fig.~\ref{fig:Diag}c. The renormalized 
Green's function $\tilde G$ is depicted as the double line, whereas single solid line corresponds to the bare dual propagator $R$.
The self-energy part $\tilde \Sigma$, by the assumption above has the following diagrammatic form (Fig.~\ref{fig:Diag}b),
and can be expressed as follows:
\begin{equation}
  \label{eq:Sigma-dual}
  \tilde{\Sigma}_{\mb{R}} = \tilde{G}^{3}_{\mb{R}}B^{2},\quad \mb{R} = \mb{r}_{i} - \mb{r}_{j}.
\end{equation}
where
\begin{equation}
  B = \left\langle\left(\frac{1}{\alpha^{-1} - \Delta}\right)^2\right\rangle - \left\langle\frac{1}{\alpha^{-1} - \Delta}\right\rangle^{2}.
\end{equation}
It is worth to mention again that the CPA could be represented as an equality between the renormalized and bare lines in dual diagrams.  

Given some $\Delta$ and $\tilde \Sigma_{\mb{R}}$ , we perform the Fourier transform of $\tilde \Sigma$ to the $k$-space, and obtain a guess for $\tilde G$ from the Dyson equation (\ref{Dyson-dual}). We transform thus obtained $\tilde G$ to the real space, substitute it into (\ref{eq:Sigma-dual}) and obtain a new  guess for $\tilde \Sigma_{\mb{R}}$. This ``inner'' iteration loop converges with some $\tilde G$ for a given $\Delta$.  Then we update $\Delta$:
\begin{equation}
  \label{eq:itt.procedure}
  \Delta_n = \Delta_{n-1} + \xi \sum_{\mb{k}}\tilde{G}(\mb{k})_{n - 1},
\end{equation}
where $\xi$ is a dimensionless factor, dedicated to improve the convergence, and repeat the inner loop.
These updates (the ``outer'' iteration loop) are repeated, until $\Delta$ is converged to the certain stable value.
Clearly, the stable point of (\ref{eq:itt.procedure}) satisfies the self-consistency condition (\ref{SCC-DVA}).  
The similar procedure can be used in the CPA calculations, but the inner loop is not needed, since $\tilde{\Sigma}^{\mathrm{CPA}}=0$.

\section{Calculations}
\label{sec:Calc}
For the model under consideration we have performed calculations of the response function using different methods, described above.
Results of the numerical simulations are used as reference to compare with the results of our analytical approaches.

Numerical simulations were performed for the square lattice formed by $N = (32\times32)$ nodes with periodic boundary conditions. 
The average response function is calculated by averaging the response functions for a set of random system configurations.
For each system configuration the response function is calculated directly. 
This leads to the series of multiple inversion of the matrices with corresponding dimension $N^{2}$, the typical number of realizations is $5\cdot2^{10}$. 
The response function for different system parameters can be calculated independently, and it is natural to perform these calculations in a parallel way.
Taking into account that at each realization of the system only the diagonal elements of the corresponding matrix are changing, a special algorithm can be used for the inversion of this set of matrices.
This algorithm allows to efficiently vectorize computations and perform the calculations using the GPGPU (general-purpose graphics processing unit). 

Calculations are organized as follows: the regular part of the matrix is inverted on the CPU using a standard numerical algorithm, 
then, for each system realization the diagonal matrix elements are changed consequently, and the inversion of the matrix is performed on the GPU, using relation (based on the Sherman-Morrison formula):
\begin{equation}
\label{GPU-inversion}
	G'_{ij}= G_{ij} - \frac{\delta M}{1 + M_{ll}\cdot\delta M}G_{il}G_{lj},
\end{equation}
where $G'_{ij}$ is inverted from matrix $M_{ij}$ with $l$-th diagonal element changed to $M'_{ll} = M_{ll} + \delta M$.
This algorithm allows to perform calculations up to 70-times faster than corresponding computations on the single CPU.

We calculate the dependence of the response function on the inverse single-particle susceptibility for systems with different filling fractions, and compared it with analytical results.
We apply the idea of the NLCPA for the system under consideration with the BZ divided into $N_c=4$ tiles (($2\times2$) -- tiling).
For systems with different filling fractions we performed calculations of the dependence of the response function on the inverse susceptibility of a single particle, and performed comparison with analytical results.
Since the susceptibility of a single particle is frequency dependent, these dependencies can be discussed as spectral dependencies.
Let us discuss the results of analytical approaches applied to the system under consideration.

\section{Results}
\label{sec:Res}
For presentation purposes, it is better to switch from the susceptibility $\alpha$ to a frequency argument.
We suppose $\alpha=\alpha_0/(\omega-\omega_0+ \imath\gamma)$ with a small loss factor $\gamma$.
\begin{figure}[h]
\includegraphics[scale = 0.70]{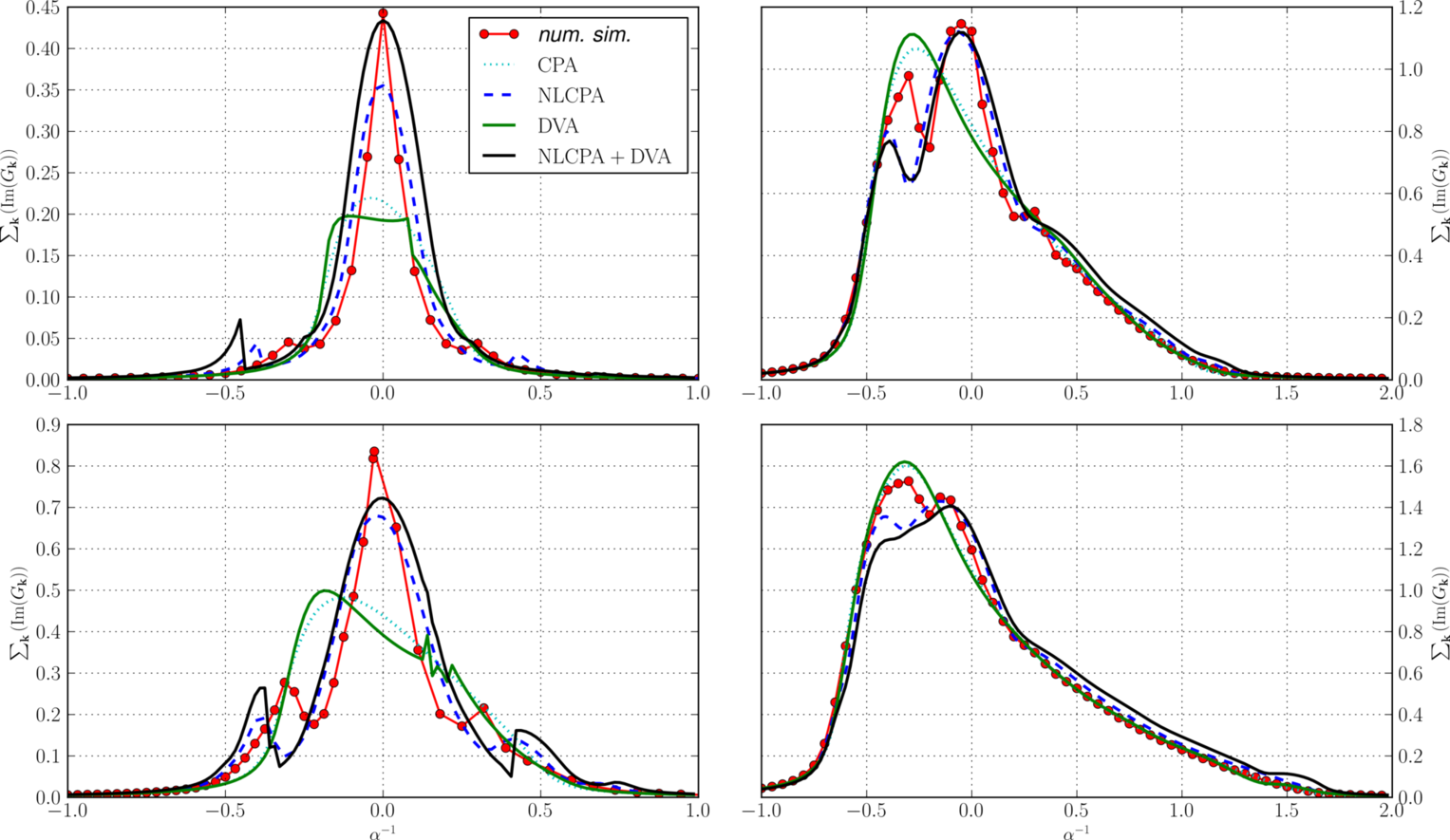}
\caption{Imaginary part of the response function for systems with different filling fractions (interaction constant $C_0=0.30$, concentration $c = 10, 20, 30, 50\%$).}
\label{fig:DOS}
\end{figure}

\begin{figure}[h]
\includegraphics{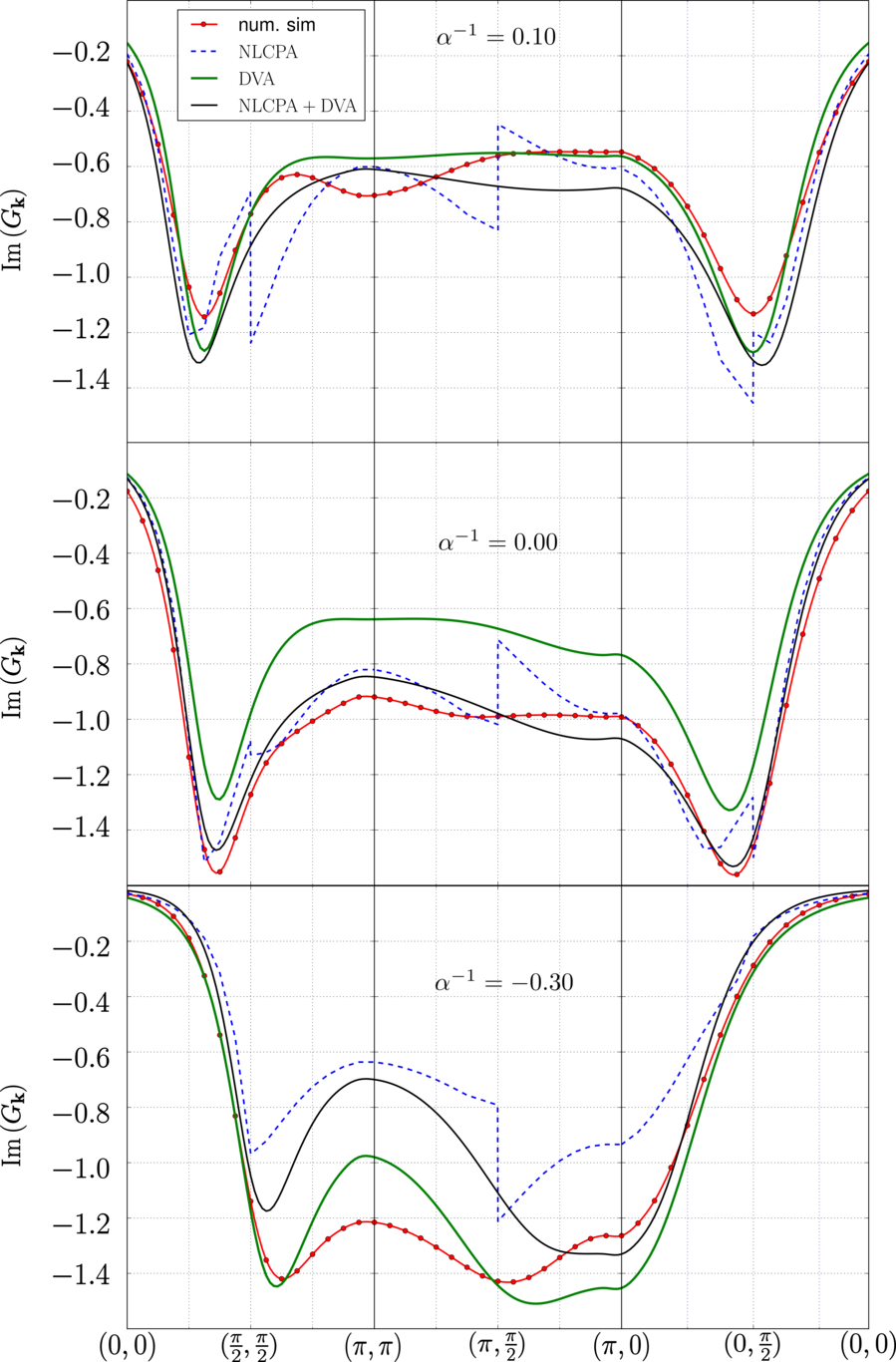}
\caption{Quasimomentum dependencies of the imaginary part of the Green function at the different $\alpha^{-1}$ points (interaction constant $C_0=0.30$, concentration $c=30\%$).}
\label{fig:Gk}
\end{figure}
Thus created spectral dependence of the density of states (that is, local part of ${\rm Im} G_{\mb{k}}$) is shown in Fig.~\ref{fig:DOS} for different filling factors. 
One can observe that the qualitative behavior of the curves is complex. 
The numerical data shows a central resonant peak and two satellites, and a broadening of the entire curve with an increase of the filling factor. As it follows from the CPA curves, the local single-particle physics is responsible for the broadening effects and the overall shape of the curve, whereas the satellite structure is completely missed in the CPA and must be therefore attributed to nonlocal effects. The displacement of the satellites from the central peak is almost independent of the filling factor. 
It is natural to conclude from this observation that the satellites appear because of certain short-range effects, like a formation of resonant dimers of two particles placed at neighboring nodes. 
Indeed, long-range collective effects involving a large number of particles would be strongly dependent on the concentration.

As it was described above, cluster methods and dual variables provide complimentary approaches to the problem of nonlocality. 
The NLCPA describes the short-range effects related to resonances due to some local configurations (for example, the formation of a resonant dimer of two neighboring particles, as discussed). 
Those effects are treated in a non-perturbative way: an infinite number of diagrams is virtually summed up.
Contrary, the DVA allows to take into account both short- and long-range correlations on equal footing, but only within a certain perturbation procedure (in our case, just taking into account the leading-order diagram). 
From this argumentation one should expect that the satellite structure is better described with the NLCPA, and indeed, our calculations confirm this. 
Similarly to the previous results for the Anderson model the NLCPA curves do show the satellites, although their positions do not always coincide with the numerical data.
On the other hand, the DVA completely misses this effect.

Now let us turn to the results for the $\mb{k}$-dependent spectral function ${\rm Im} G_\mb{k}$ near the resonance of the density of states. 
Figure~\ref{fig:Gk} shows the data exactly on the resonance and at a point on the shoulder of the curves in Fig.~\ref{fig:DOS}.
Our numerical data demonstrates a strong change of the dispersion law, that is not accurately captured by the CPA.
Besides a quantitative disagreement, at the main resonance of the density of states one can observe a qualitative difference: the numerical data along $(\pi,\pi) - (\pi,0)$ line at border of the BZ show a much flatter spectral function than the CPA predicts. This flattening can be interpreted as an effect of a strong disorder. 
It occurs for certain region of the quasimomentum values and is consequently a nonlocal effect. 

For the $\mb{k}$-dependence of the spectral function, one cannot expect a good result from the NLCPA, because the ($2\times2$) -- tiling is too crude to describe an actual $\mb{k}$-dependence, and a step-like change at the borders of tiles is non-physical.
Indeed, one can discuss only the slope of the NLCPA curves in Fig.~\ref{fig:Gk}, the detailed behavior is not reproduced.
It is also hard to speculate about a reproduction of the flattening phenomenon.
 
Contrary, the DVA describes short- and long-range phenomena at the same footing, and thus produces smooth $\mb{k}$-dependencies.
It also describes the observed flattening phenomenon, demonstrating a behavior qualitatively very similar to the numerical data. 
We conclude that the DVA handles the physics of the strong disorder near the DoS peak energy.

We should remark that such a picture is not generic for all points of the spectral curve. 
For instance, one should not expect a good result from the DVA at the satellite points, as the formation of the satellites itself is not handled by the DVA. 
Indeed, as Fig.~\ref{fig:DOS} shows, neither the NLCPA nor the DVA are satisfactory in this respect. 
In fact, the best fit of the numerical data is given by just the CPA.

More progress can be expected with use of combined schemes, including both cluster and dual-variables contributions. 
We made an effort to implement the simplest scheme of this kind:
a substitution of the short-ranged part of the self-energy calculated in the CPA by the corresponding part of the NLCPA self-energy.
Results of this method are shown in figures by the curves marked as ``\textrm{NLCPA+DVA}'', and it is seen that this scheme doesn't works satisfactory:
no significant improvement is found; the curve shows kinks and a wrong behavior at the shoulders.
A more advanced combined method is clearly needed.

To summarize, using the identity (\ref{GGdual}) we presented a very simple formulation of the dual-variables diagram expansion around the CPA result for the response of the systems with a diagonal disorder. 
Similar to the CPA itself, a formal small parameter can be explicitly pointed out in the limiting cases (e.g. small disorder or large coordination number). 
For a practically interesting situation of a strongly disordered low-dimensional system the applicability of the method is based on a physical observation that a large part of correlations is local.
In the  DVA, this local part is taken into account via the choice of new variables, by the condition (\ref{SCC-DVA}), whereas the nonlocal correlations are treated diagrammatically.
Such an approach captures long-range spatial correlations.
The scheme is complimentary to the short-range NLCPA (cluster) methods and is suitable for a description of the $\mb{k}$-dependent quantities such as the self-energy.

\begin{acknowledgments}
The work was supported by Federal Program of the Russian Ministry of the Education and Science, grant  07.514.12.4033. 
\end{acknowledgments}
\bibliographystyle{unsrt}
\bibliography{main}
\end{document}